\documentclass[reprint,showpacs,aps,pra]{revtex4-1}
\usepackage{graphics}
\usepackage{bm}
\usepackage{bbm}
\usepackage{amsmath}
\usepackage{amssymb}
\usepackage{graphicx}
\usepackage{amsthm}
\begin{document}

\title{Bipartite entanglement in  fermion systems}
\author{N.\ Gigena, R.\ Rossignoli}
\affiliation{IFLP-Departamento de F\'{\i}sica-FCE,
Universidad Nacional de La Plata, C.C. 67, La Plata (1900), Argentina}

\begin{abstract}

We discuss the relation between fermion entanglement and bipartite
entanglement. We first show that an exact correspondence between them arises
when the states are constrained to have a definite local number parity.
Moreover, for arbitrary states in a four dimensional single-particle Hilbert
space, the fermion entanglement is shown to measure the entanglement between
two distinguishable qubits defined by a suitable partition of this space. Such
entanglement can be used as a resource for tasks like quantum teleportation. On
the other hand, this fermionic entanglement provides a lower bound to the
entanglement of an arbitrary bipartition although in this case the local states
involved will generally have different number parities. Finally the fermionic
implementation of the teleportation and superdense coding protocols based on
qubits with odd and even number parity is discussed, together with the role of
the previous types of entanglement. \\
pacs{03.67.Mn, 03.65.Ud, 05.30.Fk}
\end{abstract}
\maketitle

\section{Introduction}

Entanglement is a fundamental feature of quantum mechanics, and its
quantification and characterization has been one of the main goals of quantum
information theory for the last decades \cite{A.08,H.09,E.10}. It is also at
the heart of quantum information processing \cite{NC.00}, being recognized as
the key ingredient for quantum state teleportation \cite{BB.93} and the
resource that makes some pure state based quantum algorithms exponentially
faster than their classical counterparts \cite{JL.03}.

Although entanglement has been extensively studied for systems of
distinguishable constituents, less attention has been paid to the case of a
system of indistinguishable fermions. Only in recent years the topic has gained
an increasing strength
\cite{Sch.01,SDM.01,Eck.02,Wi.03,GM.04,Ci.09,Za.02,Shi.03,Fri.13,Be.14,Pu.14,
Ci.09,Pl.09,IV.13,Os.14,Os.142,SL.14,IV.14,GR.15,M.16,GR.16}.
Mainly two different approaches may be recognized in the attempts of
generalizing the definition of entanglement to fermion systems: The first is
{\it entanglement between modes} \cite{Za.02,Shi.03,Fri.13,Be.14,Pu.14}, where
the system and subsystems consist of some collection of single-particle  modes
that can be shared. This approach requires to fix some basis of the
single-particle state space and then to specify the modes that constitute each
subsystem. The other approach is known as {\it entanglement between particles}
\cite{Sch.01,SDM.01,Eck.02,Wi.03,GM.04,Ci.09,Pl.09,IV.13,Os.14,Os.142,SL.14,IV.14,M.16},
where the indistinguishable constituents of the system are taken as subsystems
and entanglement is defined beyond symmetrization.

	In a previous work \cite{GR.15} we defined an entropic measure of mode
entanglement in fermion systems which is shown to be a measure of entanglement
between particles after an optimization over bases of the single-particle (sp)
state space is performed. Moreover, when the sp state space dimension is four
and the particle number is fixed to two, this entanglement measure reduces to
the {\it Slater correlation measure} defined in \cite{Sch.01}. In the present
work we first show that the entanglement between two distinguishable qubits is
the same as that measured by this fermionic entanglement entropy when the
fermionic states are constrained to have a fixed local number parity in the
associated bipartition of the sp space. Then we use this correspondence to show
that, in fact, any state of a fermion system with a 4-dimensional sp Hilbert
space may be seen as a state of two distinguishable qubits for a suitable
bipartition of the sp space, with its entanglement measured by the fermionic
entanglement entropy. On the other hand, for an arbitrary bipartition involving
no fixed local number parity the fermionic entanglement entropy is shown to
provide a lower bound to the associated bipartite entanglement.  As application
we use these results to show that qubit-based quantum circuits may be rewritten
as mode-based fermionic circuits if we impose the appropriate restriction to
the occupation numbers, recovering reversible classical computation when the
input states are Slater determinants (in the basis of interest). Two types of
fermionic qubit representations, based on odd or even number parity qubits, are
seen to naturally emerge. Finally, we show that the extra bipartite entanglement 
that can be obtained by relaxing this local parity restriction can in principle 
be used for protocols such as superdense coding. 

The formalism and theoretical results are provided in sec.\ \ref{II}, while
their applications are discussed in sec.\ \ref{III}. Conclusions are finally
provided in sec.\ \ref{IV}.

\section{Formalism\label{II}}

\subsection{Fermionic entanglement entropy and concurrence}

We will consider a fermion system with a single-particle (sp) Hilbert space
$\cal{H}$. We will deal with pure states $|\psi\rangle$ which do not
necessarily have a fixed particle number, although the number parity will be
fixed, in agreement with the parity superselection rule \cite{Fr.15}:  
$P|\psi\rangle=\pm|\psi\rangle$, with $P=\exp[i\pi \sum_j
c^\dagger_j c_j]$ the number parity operator. Here  $c_j$, $c^\dagger_j$ denote
fermion annihilation and creation  operators satisfying the usual
anticommutation relations
\begin{equation}
\{c_i,c_j\}=0,\;\;\{c_i,c_j^\dagger\}=\delta_{ij}\,.
\end{equation}

In \cite{GR.15} we defined a  {\it one-body entanglement entropy} for a general
pure fermion state $|\psi\rangle$,
\begin{equation}
S^{\rm sp}(|\psi\rangle)={\rm Tr}\,h(\rho^{\rm sp}),\label{Ssp}
\end{equation}
where $\rho^{\rm sp}_{ij}=\langle c^\dagger_j
c_i\rangle\equiv\langle\psi|c^\dagger_j c_i|\psi\rangle$ is the one body
density matrix of the system and $h(p)=-p\log_2 p-(1-p)\log_2(1-p)$. Eq.\
(\ref{Ssp}) is proportional to the minimum, over all sp bases of ${\cal H}$, of
the average entanglement entropy between a sp mode and its orthogonal
complement (which in turn arises from a properly defined measurement of the
occupation of a sp mode), and  vanishes iff $|\psi\rangle$ is a Slater
Determinant (SD), i.e. $|\psi\rangle=c^\dagger_1\ldots c^\dagger_k|0\rangle$.
This definition  is easily extended to quasiparticle modes, in which case
\cite{GR.15}
\begin{equation}
S^{\rm qsp}(|\psi\rangle)=-{\rm Tr}\,\rho^{\rm qsp}\log_2 (\rho^{\rm qsp})\,,\label{Sqsp}
\end{equation}
where $\rho^{\rm qsp}$ is now the extended one-body density matrix
\begin{equation}
\rho^{\rm qsp}= 1-\left\langle\left(\begin{array}{c}\bm{c}\\
\bm{c}^\dagger\end{array}\right)\left(\begin{array}{cc}\bm{c}^\dagger&\bm{c}
\end{array}\right)\right\rangle=
\left(\begin{array}{cc}\rho^{\rm sp}&\kappa\\-\bar{\kappa}\;\;&
\mathbbm{1}-\bar{\rho}^{\rm sp}\end{array}\right)\label{qsp}\,,
\end{equation}
with $\kappa_{ij}=\langle c_jc_i\rangle$,  $-\bar{\kappa}_{ij}=\langle
c^\dagger_j c^\dagger_i\rangle$ and $(\mathbbm{1}-\bar{\rho}^{\rm
sp})_{ij}=\langle c_j c^\dagger_i\rangle$. Eq.\ (\ref{Sqsp}) vanishes iff
$|\psi\rangle$ is a quasiparticle vacuum or SD and satisfies $S^{\rm
qsp}(|\psi\rangle)\leq S^{\rm sp}(|\psi\rangle)$,  with  $S^{\rm
qsp}(|\psi\rangle)=S^{\rm sp}(|\psi\rangle)$ iff $\kappa=0$.

While Eq.\ (\ref{Ssp}) is invariant under unitary transformations
$c_i\rightarrow \sum_k \bar{U}_{ki}c_k$, $UU^\dagger=I$, which lead to
$\rho^{\rm sp}\rightarrow U^\dagger\rho^{\rm sp} U$,  Eq.\ (\ref{qsp}) remains
invariant under general Bogoliubov transformations
\begin{equation}
 c_i\rightarrow a_i=\sum_k \bar{U}_{ki} c_k+V_{ki}c^\dagger_k\,, \label{anu}
\end{equation}
where matrices $U$ and $V$ satisfy $UU^\dagger +VV^\dagger={\mathbbm 1}$ and
$UV^T+VU^T=0$ in order that $\{a_i,a^\dagger_i\}$ fulfill the fermionic
anticommutation relations \cite{RS.80}. In this case $\rho^{\rm qsp}\rightarrow
W^\dagger\rho^{\rm qsp}W$, with $W=(^{U\;V}_{\bar{V}\;\bar{U}})$ a unitary
matrix. In terms of the  operators  diagonalizing $\rho^{\rm qsp}$, we then
have
\[
1-\left\langle\left(\begin{array}{c}\bm{a}\\
\bm{a}^\dagger\end{array}\right)\left(\begin{array}{cc}\bm{a}^\dagger&\bm{a} 
\end{array}\right)\right\rangle
= \left(\begin{array}{cc}f&0\\ 0&1-f\end{array}\right)\,,
\]
with $f_{kl}=f_k\delta_{kl}$ and $f_k, 1-f_k$  the eigenvalues of $\rho^{\rm
qsp}$.

For a sp space $\cal H$ of dimension 4, $\rho^{\rm qsp}$ becomes an $8\times 8$ matrix,  
and it was shown that its eigenvalues for a pure state  $|\psi\rangle$ are 
{\it fourfold degenerate} and can be written as \cite{GR.15}
\begin{equation}
f_\pm=\frac{1\pm\sqrt{1-C^2(|\psi\rangle)}}{2}\,,\label{Cf}
\end{equation}
where $C(|\psi\rangle)=2\sqrt{f_+f_-}\in[0,1]$ is called {\it fermionic concurrence},  
in analogy with that defined for two-qubits \cite{W.98}. Eq.\ (\ref{Sqsp}) becomes 
then an increasing function of $C(|\psi\rangle)$, vanishing iff the latter vanishes. 
This fermionic concurrence can also be explicitly evaluated: Writing a general even number 
parity pure state in such a space  as 
\begin{equation}|\psi\rangle=
(\alpha_0+{\textstyle\frac{1}{2}}\sum_{i,j}\alpha_{ij}c^\dagger_ic^\dagger_j+
\alpha_4 c^\dagger_1c^\dagger_2c^\dagger_3c^\dagger_4)|0\rangle\label{eq1}\,,\end{equation}
where $\alpha_{ij}=-\alpha_{ji}$, $i,j=1,\ldots,4$  and $|\alpha_0^2|+|\alpha_4^2|
+ \frac{1}{2}{\rm tr}\,\alpha^\dagger\alpha=1$, 
then $\rho^{\rm sp}=\alpha\alpha^\dagger+|\alpha_4|^2{\mathbbm 1}$,  
$\kappa=\alpha_0^*\alpha+\alpha_4 \tilde{\alpha}^*$, 
with $\tilde{\alpha}_{ij}=\frac{1}{2}\sum_{k,l}\epsilon_{ijkl}\alpha_{kl}$ 
($\epsilon_{ijkl}$ denotes the fully antisymmetric tensor) 
and it can be shown that \cite{GR.15} 
\begin{equation} C(|\psi\rangle)=2|\alpha_{12}\alpha_{34}-\alpha_{13}\alpha_{24}
+\alpha_{14}\alpha_{23}-\alpha_0\alpha_4|\,.\label{Cex1}\end{equation}
For two-fermion states ($\alpha_0=\alpha_4=0$) it reduces to the {\it Slater correlation measure} 
defined in \cite{Sch.01,Eck.02}, for which $\kappa=0$ and $f_{\pm}$ become the eigenvalues 
(two-fold degenerate) of $\rho^{\rm sp}$. An expression similar to (\ref{Cex1}) holds for an odd 
number parity state (see \cite{GR.15} and sec.\ \ref{E}). Moreover, in such sp space the concurrence 
and the associated entanglement of formation can also be explicitly determined for arbitrary mixed 
states  \cite{Sch.01,GR.15}. 

A four-dimensional sp space (which generates an eight-dimensional state space for each value of the 
parity $P$) becomes then  exactly solvable, being as well the first non-trivial dimension since for 
${\rm dim}\, {\cal H}\,\leq 3$ any definite parity pure state can be written as a SD or quasiparticle 
vacuum \cite{GR.15}, as verified from (\ref{Cex1}) ($C(|\psi\rangle)=0$ if 
one of the sp sates is left empty). 
It is also physically relevant, since it can accommodate the  basic situation of two spin $1/2$ 
fermions at two different sites or, more generally, states of  spin $1/2$ fermions occupying just  
two  orbital states, as in a double well scenario. The relevant sp space in these cases is 
${\cal H_S}\otimes {\cal H_O}$, with ${\cal H_S}$ the  spin space and  ${\cal H_O}$ the two-dimensional 
subspace spanned by the two orbital states. In particular, just four  sp states are essentially used in 
recent proposals for observing Bell-violation from single electron entanglement \cite{DB.16}. 
\subsection{Bipartite  entanglement as two-fermion entanglement}
Let us now consider a system of two distinguishable qubits prepared in a pure
state $\alpha_+|00\rangle+\alpha_-|11\rangle$, i.e.
\begin{eqnarray}
|\psi\rangle_{AB}=\alpha_+|\uparrow\rangle_A\otimes|\uparrow\rangle_B+
\alpha_-|\downarrow\rangle_A\otimes|\downarrow\rangle_B\,,\label{st1}
\end{eqnarray}
where $|\alpha_+^2|+|\alpha_-^2|=1$ and the notation indicates a possible
realization in terms of two spin $1/2$ particles located at different sites $A,B$,
with their spins aligned parallel or antiparallel to a given ($z$) axis. We can
also consider this last state  as a two-fermion state of a spin $1/2$ fermion
system with sp  space ${\cal H}={\cal H_S}\otimes{\cal H_O}$:
\begin{equation}
|\psi\rangle_f=(\alpha_+ c^\dagger_{A\uparrow}c^\dagger_{B\uparrow}+
\alpha_- c^\dagger_{A\downarrow}c^\dagger_{B\downarrow})|0\rangle\,,\label{st2}
\end{equation}
with $|0\rangle$ the fermionic vacuum. A measurement of spin ``$A$'' or ``$B$''
along $z$ can be described in the fermionic representation by the operators
$\Pi_{S\mu}=c^\dagger_{S\mu}c_{S\mu},\,S=A,B,\,\mu=\uparrow,\downarrow$, which
satisfy  $\Pi_{S\mu}^2=\Pi_{S\mu}$ and $[\Pi_{S\mu},\Pi_{S'\mu'}]=0$, with
$\sum_\mu\Pi_{S\mu}|\psi\rangle_f= |\psi\rangle_f$. Furthermore, we can
describe any ``local'' operator on  $A$ or $B$ in terms of  Pauli operators  if
we define, for $S=A,B$,
\begin{subequations}\label{ssp}
    \begin{eqnarray}
\sigma_{S x}&=&
c^\dagger_{S\uparrow}c_{S\downarrow}+c^\dagger_{S\downarrow}c_{S\uparrow}\,,\\
\sigma_{Sy}&=& -i(c^\dagger_{S\uparrow}c_{S\downarrow}-c^\dagger_{S\downarrow}c_{S\uparrow})\,,\\
\sigma_{Sz}&=&c^\dagger_{S\uparrow}c_{S\uparrow}-c^\dagger_{S\downarrow}c_{S\downarrow}\,,
    \end{eqnarray}\end{subequations}
which verify the usual  commutation relations $[\sigma_{Sj},\sigma_{S'k}]=2i\delta_{SS'}\epsilon_{jkl}$, 
($\epsilon_{jkl}$ is the antisymmetric tensor), 
with  $\sigma_{Sj}^2|\psi\rangle_f=|\psi\rangle_f$.

It is also apparent that the state (\ref{st1}) is separable iff
$\alpha_+=0$ or $\alpha_-=0$, which  is  precisely the condition  which ensures that
the state (\ref{st2}) is a SD.  
Moreover, the standard concurrence \cite{W.98}
of the state (\ref{st1}) is  {\it identical} with the  fermionic concurrence (\ref{Cex1}) of
the state (\ref{st2}):
\begin{equation}C(|\psi\rangle_{AB})=2|\alpha_+\alpha_-|=C(|\psi\rangle_f)\,,\label{Cef}
\end{equation}
with $f_{\pm}=|\alpha_{\pm}^2|$ in (\ref{Cf}).  Entangled two-qubit states  (\ref{st1})
correspond then to two-fermion states (\ref{st2}) which are not SD's, and
vice-versa.

Such correspondence remains of course valid for any bipartite two-qubit state
\begin{equation}
|\psi\rangle_{AB}=\sum_{\mu,\nu} \alpha_{\mu\nu}|\mu\rangle_A\otimes |\nu\rangle_B\,,\label{st3}
\end{equation}
 which in the fermionic representation becomes
\begin{equation} |\psi\rangle_f=\sum_{\mu,\nu}\alpha_{\mu\nu}c^{\dagger}_{A\mu}c^\dagger_{B\nu}
|0\rangle\,.
 \label{st4}\end{equation}
We now obtain $C(|\psi\rangle_{AB})=2|{\rm det}\,\alpha|=C(|\psi\rangle_f)$, according to the 
standard and fermionic (Eq.\ (\ref{Cex1})) expressions. These states can in fact be taken to 
the previous Schmidt forms (\ref{st1})-(\ref{st2})  
(with  $|\alpha_{\pm}|$ the singular values of the matrix $\alpha$) by means
of local unitary transformations, which in the fermionic representation become 
$c_{S\mu}\rightarrow \sum_{\nu}\bar{U}^S_{\nu\mu}c_{S\nu}$. 

Previous considerations remain also valid for {\it general} bipartite
states of systems of {\it arbitrary} dimension ($\mu=1,\ldots,d_A$,
$\nu=1,\ldots,d_B$ in (\ref{st3})--(\ref{st4})),  if the sp space of the associated fermionic 
system (of dimension $d_A+d_B$)  is decomposed as  ${\cal H}_A\oplus {\cal H}_B$. The sp 
density matrix $\rho^{\rm sp}$ derived 
from the state (\ref{st4}) 
 takes in the general case the blocked form
\begin{equation}
\rho^{\rm sp}=\begin{pmatrix}\alpha\alpha^\dagger&0\\0&\alpha^T\bar{\alpha}\end{pmatrix}
=\begin{pmatrix}\rho_A&0\\0&\rho_B
\end{pmatrix}\label{rsp2x}\,,
 \end{equation}
i.e.\ $\langle c^\dagger_{S\nu}c_{S'\mu}\rangle=\delta_{SS'}(\rho_S)_{\mu\nu}$,
where $\rho_{A(B)}$  are the local
reduced density matrices ${\rm Tr}_{B(A)}|\psi\rangle_{AB}\langle\psi|$ of the state (\ref{st3}) 
in the standard basis. Hence, in the fermionic setting  $\rho^{\rm sp}$
contains the information of {\it both} local states and its diagonalization implies that of 
{\it both} $\rho_A$ and $\rho_B$. Its eigenvalues will then be those of these matrices, being 
hence two-fold degenerate and equal to  the square of the singular values  of 
 the matrix $\alpha$ (becoming $f_{\pm}=|\alpha_{\pm}|^2$ in the two-qubit case).  
In the general case, the entanglement entropy of the  sate (\ref{st3}) can then be written as
\begin{equation} E(A,B)=S(\rho_A)=S(\rho_B)={\textstyle\frac{1}{2}}S(\rho^{\rm sp})\,,\label{rsf}
\end{equation}
which holds for the von Neumann entropy $S(\rho)=-{\rm Tr}\,\rho\log_2\rho$
as well as for any trace form entropy \cite{CR.02} $S(\rho)={\rm Tr}\,f(\rho)$
($f$ concave,  $f(0)=f(1)=0$). Thus, the entanglement entropy of the general bipartite state 
(\ref{st3}) is just proportional to the fermionic entanglement entropy (as defined in (\ref{Ssp})) 
of the associated state (\ref{st4}). Hence, for any dimension there is an exact correspondence 
between the bipartite states (\ref{st3})  and the
two-fermion states (\ref{st4}), with local operators represented by linear
combinations of  one-body local fermion  operators $c^\dagger_{S\nu}c_{S\mu}$
(satisfying $[c^\dagger_{A\mu}c_{A\nu},c^\dagger_{B\mu'}c_{B\nu'}]=0$) and 
$|\psi\rangle_{AB}$ entangled iff $|\psi\rangle_f$ is not a SD. 

This equivalence holds also for mixed states i.e., convex combinations of the states (\ref{st3}) 
and (\ref{st4}). The bipartite states will be separable, i.e., convex combinations of
product states \cite{WW.89} iff the associated fermionic mixed state can be written as a
convex combination of SD's of the form (\ref{st4}). In particular, for
two-qubit states a four-dimensional sp fermion space suffices and 
the standard mixed state concurrence \cite{W.98} will coincide exactly
with the fermionic mixed state concurrence 
\cite{Sch.01,Os.14,Os.142,SL.14,GR.15} of mixtures of states  (\ref{st4}). 

\subsection{Bipartite entanglement as quasiparticle fermion entanglement \label{C}}
Other fermionic representations of the state (\ref{st3}) with similar
properties are also feasible. For instance, in the two qubit case we can
perform a particle hole-transformation of the fermion operators with spin down,
\begin{equation}
c^\dagger_{S\uparrow}\longrightarrow c^\dagger_{S\uparrow},\;\;
c^\dagger_{S\downarrow}\longrightarrow  c_{S\downarrow},\;\; \;S=A,B
\end{equation}
such that the aligned state $|\downarrow\rangle_A\otimes |\downarrow\rangle_B$
corresponds now to the vacuum of the new operators ($|0\rangle\longrightarrow
c^\dagger_{A\downarrow}c^\dagger_{B\downarrow}|0\rangle$), with the new
$c^\dagger_{S\downarrow}$ creating a hole. The remaining states of the standard
basis become one and two particle-hole excitations. We can then rewrite the
state (\ref{st2}) as
\begin{eqnarray}
|\tilde{\psi}\rangle_f&=&(\alpha_{-}+
\alpha_{+}c^\dagger_{A\uparrow}c^\dagger_{A\downarrow}c^\dagger_{B\uparrow}c^\dagger_{B\downarrow})
|0\rangle\,,\label{st5}
\end{eqnarray}
i.e., as a superposition of the vacuum plus two particle-hole excitations, 
with each ``side'' having now either 0 or two fermions, i.e., an {\it even} local number parity
 ($e^{i\pi N_S}=1$ for $S=A,B$,  $N_S=\sum_{\mu}c^\dagger_{S\mu} c_{S\mu}$). It is apparent that 
 the state (\ref{st5}) is a a quasiparticle  vacuum or SD iff  $\alpha_+=0$ or $\alpha_-=0$.    
Moreover, for the state (\ref{st5}) Eq.\ (\ref{Cex1}) leads again to $C(|\tilde\psi\rangle_f)
=2|\alpha_+\alpha_-|$,  implying the equivalence (\ref{Cef}) between the bipartite and the present 
generalized fermionic concurrrence, invariant under Bogoliubov (and hence particle-hole) transformations. 
 
The  local Pauli operators (\ref{ssp}) become now 
 \begin{subequations}\label{ssp2}\begin{eqnarray}\tilde{\sigma}_{S x}&=&
 c^\dagger_{S\uparrow}c^\dagger_{S\downarrow}+c_{S\downarrow}c_{S\uparrow}\,,\\
 \tilde{\sigma}_{Sy}&=&
  -i(c^\dagger_{S\uparrow}c^\dagger_{S\downarrow}-c_{S\downarrow}c_{S\uparrow})\,,\\
 \tilde{\sigma}_{Sz}&=&
 c^\dagger_{S\uparrow}c_{S\uparrow}+c^\dagger_{S\downarrow}c_{S\downarrow}-1\,,
 \end{eqnarray}
   \end{subequations}
which  verify the same $SU(2)$ commutation relations $[\tilde{\sigma}_{Sj},\tilde{\sigma}_{S'k}]=
2i\delta_{SS'}\epsilon_{jkl}\tilde{\sigma}_{Sl}$, with  $\tilde{\sigma}_{Sj}^2|
\tilde\psi\rangle_f=|\tilde\psi\rangle_f$ $\forall j$. Any local operation  can be written in terms 
of these operators, which represent now local paticle-hole creation or annihilation  and counting. 

Similarly, we may write the general two-qubit state (\ref{st3}) as
\begin{eqnarray}
|\tilde{\psi}\rangle_f&=&\sum_{\mu,\nu}\alpha_{\mu\nu}
(c^\dagger_{A\uparrow}c^\dagger_{A\downarrow})^{n_\mu}
(c^\dagger_{B\uparrow}c^\dagger_{B\downarrow})^{n_\nu}
|0\rangle\,,\label{st55}
\end{eqnarray}
where $\mu,\nu=\pm$ and $n_{-}=0$, $n_{+}=1$. This state can be brought back to the ``Schmidt''
form (\ref{st5}) by means of  ``local'' Bogoliubov transformations 
$c_{S\uparrow}\rightarrow u_S c_{S\uparrow}+v_S c^\dagger_{S\downarrow}$,
$c_{S\downarrow}\rightarrow u_S c_{S\downarrow}-v_S c^\dagger_{S\uparrow}$,
$|u_S^2|+|v_S^2|=1$,  which will diagonalize $\rho^{\rm qsp}$ (see below) and change the vacuum 
as $|0\rangle\rightarrow [\prod_{S=A,B}(u_S-v_Sc^\dagger_{S\uparrow}c^\dagger_{S\downarrow})]|0\rangle$. 
It is again verified that for this state Eq.\ (\ref{Cex1}) leads to $C(|\tilde\psi\rangle_f)
=2|{\rm det}\,\alpha|=2|\alpha_+\alpha_-|$, with $|\alpha_{\pm}|$ the singular values of the 
matrix $\alpha$. The state (\ref{st3}) is then entangled iff the state (\ref{st55}) is not a 
{\it quasiparticle} vacuum or SD ($C(|\tilde\psi\rangle_f)>0$). 

In this case the extended density matrix $\rho^{\rm qsp}$ is to be considered,
with elements  $\langle c^\dagger_{S\nu}
c_{S'\mu}\rangle=\delta_{SS'}\delta_{\mu\nu}p_S$, $\langle
c_{S\nu}c_{S'\mu}\rangle=\delta_{SS'}\delta_{\nu,-\mu}(-1)^{n_\mu}q_S$, where
$p_{A(B)}=|\alpha_{++}|^2+|\alpha_{+-(-+)}|^2$,
$q_{A(B)}=\alpha_{++}\alpha^*_{-+(+-)}+\alpha_{+-(-+)}\alpha^*_{--}$. For the Schmidt 
form (\ref{st55}), $\rho^{\rm qsp}$ becomes diagonal (($p_{A(B)}=|\alpha_+|^2$, $q_{A(B)} = 0$). 
 Reduced states $\rho_{A(B)}$ are now to be recovered as particular blocks of $\rho^{\rm qsp}$:
\begin{eqnarray}
\rho_S&=&\frac{1}{2}\begin{pmatrix}1+\langle\tilde{\sigma}_{Sz}\rangle&\langle\tilde{\sigma}_{Sx}
\rangle-i\langle\tilde{\sigma}_{Sy}\rangle\\\langle\tilde{\sigma}_{Sx}\rangle+
i\langle\tilde{\sigma}_{Sy}\rangle &
1-\langle\tilde{\sigma}_{Sz}\rangle\end{pmatrix}\nonumber\\&=& 
\begin{pmatrix}\langle c^\dagger_{S\uparrow}c_{S\uparrow}\rangle&
\langle c_{S\downarrow}c_{S\uparrow}\rangle\\
\langle c^\dagger_{Si\uparrow}c^\dagger_{S\downarrow}\rangle&
\langle c_{S\uparrow}c^\dagger_{S\uparrow}\rangle
\end{pmatrix}\,.\label{raq}
\end{eqnarray}
Diagonalization of $\rho^{\rm qsp}$ will, nevertheless, still imply  that of $\rho_A$ 
and $\rho_B$. It is verified that its eigenvalues are $f_{\pm}=|\alpha_{\pm}|^2$,  four-fold
degenerate, with $|\alpha_{\pm}|$ the singular values of the matrix $\alpha$.
We then have 
\begin{equation}
E(A,B)=S(\rho_A)=S(\rho_B)={\textstyle\frac{1}{4}} S(\rho^{\rm qsp})\,,
 \label{Sfq}\end{equation}
again valid for any trace-form entropy $S(\rho)={\rm Tr}\,f(\rho)$. And for convex
mixtures of states of the form (\ref{st55}) (whose rank will be at most $4$), the 
mixed state fermionic concurrence, as defined in \cite{GR.15}, will again coincide 
exactly with the standard two-qubit concurrence.  

The same considerations hold for general bipartite states (\ref{st3}) of
systems of arbitrary dimension  if a particle hole transformation (or in
general, a Bogoliubov transformation) is applied to the original fermion
operators in (\ref{st4}). In such a case Eq.\  (\ref{Sfq}) is valid for
entropic functions satisfying $f(p)=f(1-p)$ (a reasonable assumption as $p$
represents an average occupation number of particle or hole), since $\rho^{\rm
qsp}$ will have eigenvalues $f_k$ and $1-f_k$, now two-fold degenerate, with $f_k$ 
those of the local states $\rho_{A(B)}$. 

A final remark is that  the representations (\ref{ssp}) and (\ref{ssp2}) of
Pauli operators can coexist independently  since
\begin{equation}[\sigma_{Sj},\tilde{\sigma}_{S'k}]=0\,,\end{equation}
$\forall$ $j,k$ for both $S'\neq S$ {\it and} $S'=S$ ($SU(2)\times SU(2)$
structure \cite{RP.85} at each side $A$ or $B$).  Moreover, the even local parity states
(\ref{st55}) belong to the kernel of the operators (\ref{ssp}), while the odd
local parity states (\ref{st4}) ($e^{i \pi N_S}=-1$)  belong to the kernel of
the operators (\ref{ssp2}):
   \begin{equation}
\sigma_{Sj}|\tilde{\psi}\rangle_f=\tilde{\sigma}_{Sj}|\psi\rangle_f=0\,,\label{kern}
\end{equation}
for $S=A,B$ and $j=x,y,z$. Hence, unitary operators $e^{i\sum_{j}
\lambda_j\sigma_{Sj}}$ ($e^{i\sum_{j} \lambda_j  \tilde{\sigma}_{Sj}}$) will
become identities when applied to states $|\tilde{\psi}\rangle_f$
($|\psi\rangle_f$). A fermion system with a sp space of dimension $4$ can then
accommodate two distinct two-qubit systems, one for each value of the local
number parity, keeping the total number parity fixed ($e^{i\pi(N_A+N_B)}=1$).

\subsection{Bipartite entanglement with no fermion entanglement}
Previous examples show an exact correspondence between bipartite and fermion
entanglement. The representations considered involve not only a fixed value
of the global parity, but also of the {\it local} number parity. It is
apparent, however, that it is also possible to obtain bipartite entanglement
from SD's by choosing appropriate partitions of the sp space, although in this
case the local parity will not be fixed. For instance, the single fermion 
state 
\begin{equation}
|\psi\rangle_f=(\alpha c^\dagger_{A\uparrow}+\beta
c^\dagger_{B\uparrow})|0\rangle\,,
 \label{psis}\end{equation}
where the fermion is created in a state with no definite position if $\alpha\beta\neq 0$, 
leads  obviously to $S(\rho^{\rm sp})=0$ but corresponds to an entangled state 
$\alpha |\uparrow\rangle_A\otimes|0\rangle_B+\beta|0\rangle_A\otimes |\downarrow\rangle_B$. 
However, the local states at each side have different number parity. The same occurs with the 
two-fermion SD $(\alpha c^\dagger_{A\uparrow}+\beta c^\dagger_{B\uparrow})
(\alpha' c^\dagger_{A\downarrow}
+\beta 'c^\dagger_{B\downarrow})|0\rangle$, which has zero fermionic concurrence but 
corresponds to the entangled state $\alpha\beta'|\uparrow\rangle_A\otimes |\downarrow\rangle_B
-\alpha'\beta|\downarrow\rangle_A\otimes |\uparrow\rangle_B
+\alpha\alpha'|\uparrow\downarrow\rangle_A\otimes |0\rangle_B+\beta\beta'|0\rangle_A\otimes
|\uparrow\downarrow\rangle_B$. 

Hence, although there is entanglement with respect to the $(A,B)$ partition,  it is not possible 
to make arbitrary linear combinations of the eigenstates of $\rho_A$ or $\rho_B$, since they may 
not have a definite number parity. While  such entanglement may be sufficient for observing Bell 
inequalities violation, as proposed in \cite{DB.16}, it can exhibit  limitations for other tasks 
involving superpositions of local eigenstates, as discussed in sec.\  \ref{III}. This effect will 
occur whenever one of the fermions is created in a state which is ``split'' by the chosen 
partition of the sp space. With the restriction of a fixed number parity at each ``side'' 
an equivalence between bipartite and fermionic entanglement can become feasible, as discussed next. 
Notice that such restriction directly implies  {\it blocked} sp density matrices  $\rho^{\rm sp}$ 
and $\rho^{\rm qsp}$, since all contractions $\langle c^\dagger_{Ai}c_{Bj}\rangle$ and 
$\langle c^\dagger_{Ai}c^\dagger _{Bj}\rangle$ linking both sides do not conserve the local parity 
and will therefore {\it vanish} $\forall$ $i,j$. 

\subsection{Fermion entanglement as two-qubit entanglement \label{E}}
Let us now return to the two-fermion state  (\ref{st2}).  The reason why the two
particles become distinguishable is that the ``position'' observable allows us
to split the sp state space $\cal H$ as the direct sum of two copies of the
spin space ${\cal H_S}$, ${\cal H}={\cal H}_{{\cal S}_A}\oplus {\cal H}_{{\cal
S}_B}$,  with $_A\langle \mu|\mu\rangle_B=\langle 0|
c_{A\mu}c^\dagger_{B\mu}|0\rangle=0$ for $\mu=\uparrow$ or $\downarrow$. 
This last condition ensures in fact that there is just one fermion at each side 
($N_{A(B)}|\psi\rangle_f=|\psi\rangle_f$). However, for a more general two-fermion 
state, like that considered in the previous section, it is no longer possible to 
perform a measurement of the spin of only one particle by coupling it with position 
since both particles may be found at the same site. 

But now nothing prevents us from turning back the argument and state that if
for an arbitrary state $|\psi\rangle_f$, it is possible to split ${\cal H}$ as
${\cal H}={\cal H}_{A}\oplus{\cal H}_{B}$, where ${\cal H}_{A}$ and ${\cal
H}_{B}$ contain just one fermion ($N_A|\psi\rangle_f=N_B|\psi\rangle_f=1$),
then we recover again a system of two distinguishable qubits. 
This last feature leads us to the following important result: 

{\it Lemma 1}:
Let $|\psi\rangle_f$ be an arbitrary pure state of a fermion system with a
4-dimensional sp space $\cal H$, having definite number parity yet not
necessarily fixed fermion number. Then the entropy (\ref{Sqsp}) of the
corresponding density matrix $\rho^{\rm qsp}$ is proportional to the
entanglement entropy between the two distinguishable qubits that can be
extracted just by measuring the appropriate observables.
\begin{proof}
We start with a general state $|\psi\rangle_f$ with even number parity, which 
in this space will have the form (\ref{eq1}). For general  $\alpha_{ij}$, $\alpha_0$ 
and $\alpha_4$ in (\ref{eq1}), the basis of the sp space $\cal H$ determined by the 
fermion operators $\{c_i,c^\dagger_i\}$ cannot be split in order to measure only one 
particle at each part. This fact remains true even if $\alpha_0=\alpha_4=0$, 
as $\alpha$ is a general antisymmetric matrix. However, as proved in \cite{GR.15}, 
it is always possible to find another basis of $\cal H$, determined by fermion operators
$\{a_i,a^\dagger_i\}$ related to $\{c_i,c^\dagger_i\}$  through a Bogoliubov
transformation, such that the state (\ref{eq1}) can be rewritten as
\begin{equation}
|\psi\rangle_f=(\alpha_+a^\dagger_1 a^\dagger_2+\alpha_-
 a^\dagger_3a^\dagger_4)|0\rangle\label{sd2}\,,\end{equation}
which is analogous to Eq.\ (\ref{st2}). Here $|\alpha_{\pm}|^2=f_{\pm}$ are
just the distinct eigenvalues  (\ref{Cf})  of the extended
density matrix  $\rho^{\rm qsp}$ determined by the state
(\ref{eq1}), whereas $\{a_i,a^\dagger_i\}$ are suitable quasiparticle operators 
diagonalizing $\rho^{\rm qsp}$. The concurrence (\ref{Cex1}) becomes 
$C(|\psi\rangle_f)=2|\alpha_+\alpha_-|$. 

We then recognize (\ref{sd2}) as the Schmidt decomposition (\ref{st1}) of a
two-qubit state written in the fermionic representation (\ref{st2}),  since,
for instance, the sets $\{a^\dagger_1, a^\dagger_3\}$  and $\{a^\dagger_2,
a^\dagger_4\}$ (analogous to
$\{a^\dagger_{A\uparrow},a^\dagger_{A\downarrow}\}$ and
$\{a^\dagger_{B\uparrow},a^\dagger_{B\downarrow}\}$) span subspaces ${\cal
H}_{A}$ and ${\cal H}_{B}$ with $N_A=N_B=1$
($N_{A(B)}|\psi\rangle_f=|\psi\rangle_f$).  And because the Schmidt
coefficients $|\alpha_{\pm}|^2$ coincide with the eigenvalues of $\rho^{\rm
qsp}$, we obtain again $S(\rho_A)=S(\rho_B)=\frac{1}{4}S(\rho^{\rm qsp})$ (Eq.\
(\ref{Sfq})), with the fermionic concurrence coinciding exactly with the standard one.  

The case of general odd parity states, which in this sp space are linear 
combinations of states with one and three fermions,
\begin{equation}
|\psi\rangle_f=\sum_{i=1}^4\beta_i c^\dagger_i|0\rangle+
 {\tilde\beta}_i c_i|\bar{0}\rangle\,,\label{odd}\end{equation}
where $|\bar{0}\rangle=c^\dagger_1c^\dagger_2c^\dagger_3c^\dagger_4|0\rangle$ and 
$c_i|\bar{0}\rangle={\textstyle\frac{1}{3!}}
\sum_{j,k,l}\epsilon_{ijkl}c^\dagger_jc^\dagger_kc^\dagger_l|0\rangle$, can be
treated in a similar way, as they can be converted to even parity states of
the form (\ref{eq1}) by a particle-hole transformation of one of the states (i.e., 
$c^\dagger_1\rightarrow c_1$, $|0\rangle\rightarrow c^\dagger_1|0\rangle$, leading to
$\alpha_0=\beta_1$, $\alpha_4=-\tilde{\beta}_1$, $\alpha_{1j}=-\beta_j$,
and $\alpha_{ij}=\sum_k\epsilon_{ijk1}
\tilde{\beta}_{k}$ for $i,j=2,3,4$ in Eq.\ 
(\ref{eq1})). They can then be also written in the form (\ref{sd2}), in terms of 
suitable quasiparticle operators diagonalizing $\rho^{\rm qsp}$, so that the previous 
considerations still hold. The concurrence of the states (\ref{odd}), given by \cite{GR.15} 
$C(|\psi\rangle_f)=2|\sum_{i=1}^4 \beta_i\tilde{\beta}_i|$, becomes again 
$2|\alpha_+\alpha_-|$.  
\end{proof}

Some further comments are here in order. First, just the subspaces of $\cal H$
generated by $\{a^\dagger_1, a^\dagger_2\}$ and $\{a^\dagger_3, a^\dagger_4\}$
are defined by (\ref{sd2}), since any unitary transformation
$a^\dagger_{1(2)}\rightarrow \sum_{k=1,2}U_{k,1(2)}a^\dagger_k$ (and similarly
for $a^\dagger_{3(4)}$) will leave it  unchanged (except for phases in
$\alpha_{\pm}$).

Secondly, we may also reinterpret the state (\ref{sd2}) as a two-fermion state
with {\it even} local number parity if side $A$ is identified with operators
$\{a^\dagger_1,a^\dagger_2$\} and $B$ with $\{a^\dagger_3,a^\dagger_4$\}, such
that each side has either $0$ or two fermions ({\it even number parity
qubits}). Still with  even local number parity we may as well rewrite it in the
form (\ref{st5}),  i.e.,
\begin{equation}
|\psi\rangle_f=(\alpha_- +\alpha_+
\,a^\dagger_1a^\dagger_3a^\dagger_2a^\dagger_4)
 |0\rangle\label{sd3}\,,\end{equation}
through a transformation $a^\dagger_i\rightarrow a_i$ for $i=3,4$, with
$|0\rangle\rightarrow a^\dagger_3 a^\dagger_4|0\rangle$. Here just the vacuum
$|0\rangle$ and the completely occupied state $|\bar{0}\rangle$ are defined,
since (\ref{sd3}) remains invariant (up to a phase in $\alpha_+$)  by any
unitary transformation $a^\dagger_i\rightarrow \sum_{k}U_{ki}a^\dagger_{k}$ of
the operators $a^\dagger_i$.

Finally, if  $|\psi\rangle_f$ is a two-fermion state
$\frac{1}{2}\sum_{ij}\alpha_{ij}c^\dagger_i c^\dagger_j|0\rangle$, 
 the previous considerations remain
valid for a sp space ${\cal H}$ of arbitrary dimension. In this case $\kappa=0$
and it is always possible to rewrite $|\psi\rangle_f$ as \cite{Sch.01}
\[|\psi\rangle_f={\textstyle\sum_{k}} \alpha_k a^\dagger_{k}a^\dagger_{\bar{k}}|0\rangle\,,\]
where $|\alpha_k^2|$ are the eigenvalues of $\rho^{\rm
sp}=\alpha\alpha^\dagger$ and $\{a_k,a_{\bar{k}}\}$ are suitable fermion
operators diagonalizing this matrix, obtained through a unitary transformation
$a_{k(\bar k)}=\sum_i \bar{U}_{ik(\bar k)} c_i$  (satisfying
\cite{Sch.01}  $U^\dagger \alpha\bar{U}=\alpha'$ with $\alpha'$  a block
diagonal matrix with  $2\times 2$ blocks
$\alpha_{k}${\small$\begin{pmatrix}0&1\\-1&0 \end{pmatrix}$}). The sp space can
then be written as ${\cal H}_A\oplus{\cal H}_{B}$ with ${\cal H}_{A(B)}$ the
subspaces spanned by the sets $\{a^\dagger_{k(\bar{k})}\}$, containing each one
fermion.  We thus obtain  $S(\rho_A)=S(\rho_B)=\frac{1}{2}S(\rho^{\rm sp})$
(Eq.\ (\ref{rsf})).

\subsection{Fermion entanglement as minimum bipartite entanglement\label{F}}

We now demonstrate a second general result, concerning the mode entanglement
associated with general  decompositions  ${\cal H}={\cal H_A}\oplus {\cal H_B}$  
of a four-dimensional sp space. Any many-fermion state can be written as 
$|\psi\rangle_f=\sum_{\mu,\nu}\alpha_{\mu\nu}|\mu\nu\rangle$, where $\mu(\nu)$ 
labels orthogonal SD's on ${\cal H_A}$ (${\cal H_B}$) and $|\mu\nu\rangle=
[\prod_{i\in {\cal H_A}} (c^\dagger_i)^{n_i^\mu}]\,[\prod_{j\in{\cal H_B}}
(c^\dagger_j)^{n_j^\nu}]\,|0\rangle$ is a SD on ${\cal H}$,  with $n_i^\mu=0,1$ 
the occupation of sp state $i$ in the state $\mu$. The ensuing reduced states 
$\rho_A=\sum_{\mu,\mu'}(\alpha\alpha^\dagger)_{\mu\mu'}|\mu\rangle\langle\mu'|$ and $\rho_B=\sum_{\nu,\nu'}(\alpha^T\bar{\alpha})_{\nu\nu'}|\nu\rangle\langle\nu'|$ 
satisfy ${\rm Tr}\,\rho_{A(B)}\,O_{A(B)}={_f\langle \psi|O_{A(B)}|\psi\rangle_f}$ 
for any operator depending just on the local fermions $\{c_i,c^\dagger_i,\; 
i\in{\cal H}_{A(B)}\}$. The entanglement entropy associated with such bipartition 
is then \cite{GR.16} $E(A,B)=S(\rho_A)=S(\rho_B)$. 
 
In the present case we may have either  $2+2$ bipartitions (${\rm dim}\,{\cal H_A}
={\rm dim}\,{\cal H_B}=2$), or  $1+3$ bipartitions (${\rm dim}\,{\cal H_A}=1$, 
${\rm dim}\,{\cal H_B}=3$). In the latter the entanglement is determined just by 
the average occupation of the single state of ${\cal H_A}$ \cite{GR.15}  and corresponds 
to the case where $A$ has access to just one of the sp states possibly occupied in 
$|\psi\rangle_f$. A realization of a $2+2$ partition is just that 
of spin $1/2$ fermions which can be at two-different sites (one accessible to Alice 
and the other to Bob), while a $1+3$ bipartition could be one where  Alice has access 
to one site and just one spin direction, i.e., to the knowledge of the occupation of 
the sp state $A_{\uparrow}$. It could also apply to any asymmetric situation like that 
where spins are all up (i.e., aligned along the field direction) but the fermions can be 
in four different locations or orbital states, with only one accessible to Alice.  

{\it Lemma 2}: Let $|\psi\rangle_f$ be a general definite number parity fermion
state in a sp space ${\cal H}$ of dimension $4$, and let ${\cal H}={\cal
H}_A\oplus{\cal H}_B$ be an arbitrary decomposition of ${\cal H}$ with ${\cal
H}_A$ and ${\cal H}_B$ of finite dimension. The entanglement entropy associated
with such bipartition  satisfies
\begin{equation} S(\rho_A)=S(\rho_B)\geq {\textstyle\frac{1}{4}}S(\rho^{\rm qsp})\,.
\label{des}
\end{equation}
Eq.\ (\ref{des}) holds for {\it any} entropic form $S(\rho)={\rm Tr}f(\rho)$
($f$ concave, $f(0)=f(1)=0$). 

Hence, the fermionic entanglement represents the
{\it minimum} bipartite entanglement that can be obtained in such a space,
which is reached for those bipartitions arising from the normal forms
(\ref{sd2}) or (\ref{sd3}). The greater entanglement in a $2+2$ bipartition 
is obtained at the expense of loosing a fixed number parity in the local reduced states.
Note that $S(\rho^{\rm qsp})$ vanishes only if $|\psi\rangle_f$ is a quasiparticle 
vacuum or SD in some sp basis, while $S(\rho_{A(B)})$ does so only when the previous 
condition holds in a basis compatible with the chosen bipartition. 

\begin{figure}[t]
	\vspace*{-1.cm}
\centerline{\hspace*{-0.2cm}\scalebox{0.5}{\includegraphics{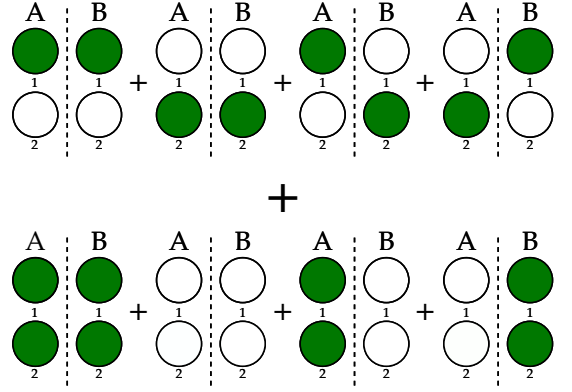}}}
\caption{Depiction of the eight even number parity  fermion states of four
single-particle modes, partitioned such that the two modes on the left of the
dashed line belong to Alice, whereas the two modes on the right to Bob. The
upper row are the states with odd local number parity (one fermion for Alice
and one fermion for Bob) whereas the bottom row those with even local number
parity (Alice and Bob may have 0 or two fermions). The states of the
bottom row can be formally obtained from those on the top by performing
particle-hole transformations $c^\dagger_{A2}\leftrightarrow c_{A2}$ and
$c^\dagger_{B2}\leftrightarrow c_{B2}$.}
	\label{fig1}
\end{figure}

We will actually show the equivalent majorization \cite{Bh.97} relation 
\begin{equation}
\lambda(\rho_{A(B)})\prec (f_+,f_-)\label{maj}\,,
\end{equation}
where $\lambda(\rho_{A(B)})$ denotes the spectrum of $\rho_{A}$ or $\rho_B$
sorted in decreasing order and $f_+$,  $f_-=1-f_+\leq f_+$ are the distinct
eigenvalues (\ref{Cf}) (fourfold degenerate)  of $\rho^{\rm qsp}$. Eq.\
(\ref{maj}) is then equivalent to the condition $\lambda_{\rm max}\leq f_+$, with 
$\lambda_{\rm max}$  the largest eigenvalue of $\rho_{A(B)}$, and  implies
(\ref{des}), while (\ref{des}) implies (\ref{maj}) if valid for any entropic
function  $f$ \cite{RC.03}.
\begin{proof}
Consider first a general even parity state (\ref{eq1}) and a $2+2$ decomposition 
${\cal H}={\cal H}_{A}\oplus {\cal H}_{B}$, with ${\cal H}_A\equiv {\cal H}_{12}$, 
${\cal H}_B\equiv {\cal H}_{34}$ and ${\cal H}_{ij}$ the subspace
generated by $\{c^\dagger_i, c^\dagger_j\}$.  Changing to the notation
$A_1, A_2,B_1,B_2$ for sp states $1,2,3,4$, we can rewrite (\ref{eq1}) 
as a sum of states of the form (\ref{st4}) and (\ref{st55}) (Fig.\ \ref{fig1}):
\begin{equation}|\psi\rangle_f
=\sum_{\mu,\nu}\beta_{\mu\nu}c^\dagger_{A_\mu}c^\dagger_{B_\nu}|0\rangle+
\sum_{\mu,\nu}\tilde{\beta}_{\mu\nu}
(c^\dagger_{A_1}c^\dagger_{A_2})^{n_\mu}(c^\dagger_{B_1}c^\dagger_{B_2})^{n_{\nu}}
|0\rangle\,,\label{Psi}\end{equation}
where  $\mu,\nu=1,2$, $\beta_{\mu\nu}=\alpha_{\mu,\nu+2}$, 
$n_\mu=\mu-1$, $\tilde{\beta}_{11}=\alpha_{0}$,
$\tilde{\beta}_{22}=\alpha_{4}$, $\tilde{\beta}_{12}=\alpha_{34}$, and
$\tilde{\beta}_{21}=\alpha_{12}$.  The first (second) sum in
(\ref{Psi}) is the odd (even) local number parity component.

After local unitary transformations $c_{S\mu}\rightarrow \sum_{\nu}
\bar{U}^S_{\nu\mu}c_{S\nu}$, $S=A,B$, which will not affect the vacuum nor the
even local parity component (except for phases in $\tilde{\beta}_{\mu\nu}$,
determined by ${\rm det}\,U^S$), we can set $\beta_{\mu\nu}$ diagonal.
Similarly, after local Bogoliubov transformations $c_{S_1}\rightarrow u_S
c_{S_1}+v_S c^\dagger_{S_2}$, $c_{S_2}\rightarrow u_S c_{S_2}-v_S
c^\dagger_{S_1}$, $|u_S^2|+|v_S^2|=1$, with
$|0\rangle\rightarrow[\prod_{S=A,B}(u_S-v_Sc^\dagger_{S_1}c^\dagger_{S_2})]|0\rangle$,
we can set $\tilde{\beta}_{\mu\nu}$ diagonal as discussed in sec.\ \ref{C}.
Though modifying the vacuum, they will not change the form of the odd local parity
component except for  phases in  $\beta_{\mu\nu}$. Thus,
by local transformations it is possible to rewrite (\ref{Psi}) as
\begin{equation}|\psi\rangle_f=(\beta_1 c^\dagger_{A_1}c^\dagger_{B_1}+
\beta_2c^\dagger_{A_2}c^\dagger_{B_2}+\tilde{\beta}_{1}+\tilde{\beta}_2c^\dagger_{A_1}
c^\dagger_{A_2}c^\dagger_{B_1}c^\dagger_{B_2})|0\rangle\,,\label{Psi2}\end{equation}
where $|\beta_\mu|$ and $|\tilde{\beta}_{\mu}|$ are the singular values of the
$2\times 2$ matrices $\beta$ and $\tilde{\beta}$ in (\ref{Psi}). Eq.\
(\ref{Psi2}) is the Schmidt decomposition for this partition, with
$(|\beta_1^2|,|\beta_2^2|, |\tilde{\beta}_1^2|$, $|\tilde{\beta}_2^2|$) the
eigenvalues of the reduced density matrices $\rho_A$ and $\rho_B$ of  modes 
$(A_1,A_2)$ and $(B_1,B_2)$ respectively. 

Now, suppose $\lambda_{\rm max}=|\beta_1^2|$.  We have
    \begin{equation}
|\beta_1|^2\leq |\beta_1|^2+|\tilde{\beta}_2|^2=\langle c^\dagger_{A_1}c_{A_1}\rangle\,.
    \label{desb1}   \end{equation}
But $\langle c^\dagger_{A_1}c_{A_1}\rangle=\sum_{k=1}^8|W_{A_1,k}|^2f_k$, where
$f_k$ are the eigenvalues of $\rho^{\rm qsp}$ (equal to $f_+$ or $f_-$)  and
$W$ the unitary matrix diagonalizing $\rho^{\rm qsp}$
($\sum_{k=1}^8|W_{A_1,k}|^2=1$).  Therefore,
\begin{equation} f_-\leq \langle c^\dagger_{A_1}c_{A_1}\rangle\leq f_+\,.\label{desb2}
\end{equation}
Eqs.\ (\ref{desb1})--(\ref{desb2}) imply $|\beta_1|^2\leq f_+$,
which demonstrates Eq.\ (\ref{maj}) and hence (\ref{des}) for a general $2+2$
bipartition ${\cal H}_{A}\oplus {\cal H}_B$.   For  $\lambda_{\rm max}$ equal
to any other coefficient  the proof  is similar.

Moreover, Eq.\ (\ref{desb2}) also shows that the sorted spectrum
$\lambda(\rho_{A_1(A_2,B)})=(\langle c^\dagger_{A_1}c_{A_1}\rangle,1-\langle
c^\dagger_{A_1}c_{A_1}\rangle)^{\downarrow}$ associated with the $1+3$
bipartition ${\cal H}_{A_1}\oplus{\cal H}_{A_2,B}$ satisfies
$\lambda(\rho_{A_1,(A_2,B)})\prec (f_+,f_-)$. In the latter $S(\rho_{A_1})$ is
the entanglement between the sp mode $A_1$ and its orthogonal complement as
defined in \cite{GR.15,GR.16}, determined by the average occupation $\langle
c^\dagger_{A_1}c_{A_1}\rangle$ of the mode. Hence,  Eqs.\
(\ref{des})--(\ref{maj}) hold as well for any $1+3$ bipartition.

And equality in (\ref{des}) for all entropic functions is evidently reached
only for those bipartitions arising from  the normal forms
(\ref{sd2})--(\ref{sd3}): Considering the non-trivial case $f_+<1$, if equality
in (\ref{des}) is to hold for all entropies, necessarily $\rho_{A(B)}$ should
be of rank $2$ with $\lambda(\rho_{A(B)})=(f_+,f_-)$.  For a $1+3$ bipartition,
this identity directly implies $\langle c^\dagger_{A_1}c_{A_1}\rangle=f_+$ or
$f_-$ and hence a bipartition arising from a normal form
(\ref{sd2})--(\ref{sd3}), where $A\equiv A_1$ is one of the sp states of the
normal basis.  And for a $2+2$  bipartition, it implies that the two eigenstates of
$\rho_A$ with non-zero eigenvalues $f_{\pm}$  should have the same number
parity, since otherwise Eq.\ (\ref{Cex1}) would imply  $C(|\psi\rangle_f)=0$
and therefore $f_+=1$, in contrast with the assumption. Hence such bipartition
must arise from a normal form (\ref{sd2}) or (\ref{sd3}).

The demonstration of previous results for odd global number parity states is
similar, as they can be rewritten as even parity states after a particle-hole
transformation.
\end{proof}

Some further comments are also in order. We may rewrite the state (\ref{Psi2}) as
\begin{equation}
|\psi\rangle_f=\sqrt{p_-}|\psi_-\rangle_f+\sqrt{p_+}|\psi_+\rangle_f\,,\label{st6}
\end{equation}
where $|\psi_{-}\rangle_f=\frac{1}{\sqrt{p_-}}(\beta_1
c^\dagger_{A_1}c^\dagger_{B_1}+\beta_2c^\dagger_{A_2}c^\dagger_{B_2})|0\rangle$,
$|\psi_+\rangle_f=\frac{1}{\sqrt{p_+}}(\tilde{\beta}_{1}+\tilde{\beta}_2c^\dagger_{A_1}
c^\dagger_{A_2}c^\dagger_{B_1}c^\dagger_{B_2})
|0\rangle$ are the normalized odd and even local parity components  and
$p_-=|\beta_1^2|+|\beta_2^2|$,
$p_+=|\tilde{\beta}_1^2|+|\tilde{\beta}_2^2|=1-p_-$. We then see that for the
von Neumann entropy, we obtain
\begin{equation}S(\rho_A)=S(\rho_B)=p_-S(\rho_A^-)+p_+S(\rho_A^+)+S(p)\,,
\end{equation}
where the first two terms represent the average of the entanglement entropies
of the odd and even local parity components
($S(\rho_A^-)=-\sum_{\mu}\frac{|\beta_\mu^2|}{p_-}\log_2\frac{|\beta_\mu^2|}{p_-}$,
$S(\rho_A^+)=-\sum_{\mu}\frac{|\tilde{\beta}_\mu^2|}{p_+}
\log_2\frac{|\tilde{\beta}_\mu^2|}{p_+}$)
while $S(p)=-\sum_{\nu=\pm} p_\nu\log_2 p_\nu$ is the  {additional} entropy
arising from the mixture of both local parities. We then have $0\leq
S(\rho_{A})\leq 2$, with the maximum $S(\rho_{A})=2$ reached iff
$S(\rho_A^\pm)=1$ and $p_{\pm}=\frac{1}{2}$.

On the other hand, the fermionic concurrence (\ref{Cex1}) of the state
(\ref{Psi2}) is just
\begin{equation}
C(|\psi\rangle_f)=2|\beta_1\beta_2+\tilde{\beta}_1\tilde{\beta}_2|\,.\label{Cs}
\end{equation}
It then satisfies
\begin{equation}
|p_-C_--p_+C_+|\leq C(|\psi\rangle_f)\leq p_-C_-+p_+C_+\,,\label{Cav}
\end{equation}
where
$C_{\pm}=C(|\psi_{\pm}\rangle_f)=2(_{|\beta_1\beta_2|/p_-}^{|\tilde{\beta}_1
\tilde{\beta}_2|/p_+})$ are the concurrences of the even and odd local parity
components. We then see, for instance, that for  {\it maximum} bipartite
entanglement $S(\rho_A)=2$, $C_{\pm}=1$ and hence $C(|\psi\rangle_f)$ can take
any value between $0$ and $1$, according to the relative phase between the even
and odd local parity components.

Finally, it is obviously possible to rewrite the Schmidt form (\ref{Psi2}) as a
two-fermion state by means of suitable local particle-hole transformations
(i.e. $c_{B_\mu}\rightarrow c^\dagger_{B_\mu}$, $\mu=1,2$, with
$|0\rangle\rightarrow c^\dagger_{B_1}c^\dagger_{B_2}|0\rangle$). After some
relabelling, we  obtain the equivalent form
\begin{equation}|\psi\rangle_f=(\beta_1c^\dagger_{A_1}c^\dagger_{B_1}+
\beta_2c^\dagger_{A_2}c^\dagger_{B_2}+
\tilde{\beta}_{2}c^\dagger_{A_1}c^\dagger_{A_2}-
\tilde{\beta}_1c^\dagger_{B_1}c^\dagger_{B_2})|0\rangle\,,\label{Psi3}
\end{equation}
where terms with two fermions at the same side side are added to the form
(\ref{st2}). Therefore, all previous considerations (\ref{st6})--(\ref{Cav})
can be realized with a fixed total number of fermions, with expression
(\ref{Cs}) still valid.

\section{Application\label{III}}
The formalism of the previous sections may now be used to rewrite a qubit-based
quantum circuit as a circuit based on fermionic modes. It is easy to see by now
that any pair of fermionic modes, say $i,j$, prepared in such a way that their
total occupation is constrained to $ N_{ij}=c^\dagger_ic_i+c^\dagger_jc_j=1$,
is essentially a qubit. Therefore, a collection of $n$  such pairs of modes
constitutes a system of $n$ qubits. Furthermore any single-qubit operation can
be performed on each pair just by using unitaries in $\cal H$ linking only
these two modes, and these unitaries can be always written in terms of the
effective Pauli operators (\ref{ssp}), i.e.,
$\sigma^{ij}_x=c^\dagger_ic_j+c^\dagger_jc_i$,
$\sigma^{ij}_y=i(c^\dagger_jc_i-c^\dagger_ic_j)$,
$\sigma^{ij}_z=c^\dagger_ic_i-c^\dagger_jc_j$. The last ingredient for
universal computation is the CNOT gate, which in the tensor product space
$A\otimes B$  can be written as $U_{_{\rm CNOT}}=|0\rangle\langle 0|\otimes
I+|1\rangle\langle 1|\otimes
\sigma_x=\exp[i\frac{\pi}{4}(1-\sigma_z)\otimes(1-\sigma_x)]$. In the fermionic
representation, if $A$ is spanned by modes $ij$ and $B$  by the different modes
$kl$, for states having one fermion at each pair of modes it can  be written as
\begin{equation}
U^f_{_{\rm CNOT}}=\exp[
i\frac{\pi}{4}(1-\sigma_z^{ij})(1-\sigma_x^{kl})]\,.\label{cnot}
\end{equation}
Since just an even number of fermion operators $c$ per pair are involved, its
action is not affected by the state of intermediate pairs. It is then possible
to implement any qubit-based quantum circuit using fermion states. 
 
\begin{figure}[t]
\vspace*{-1.cm}
\centerline{\hspace*{-0.2cm}\scalebox{.5}{\includegraphics{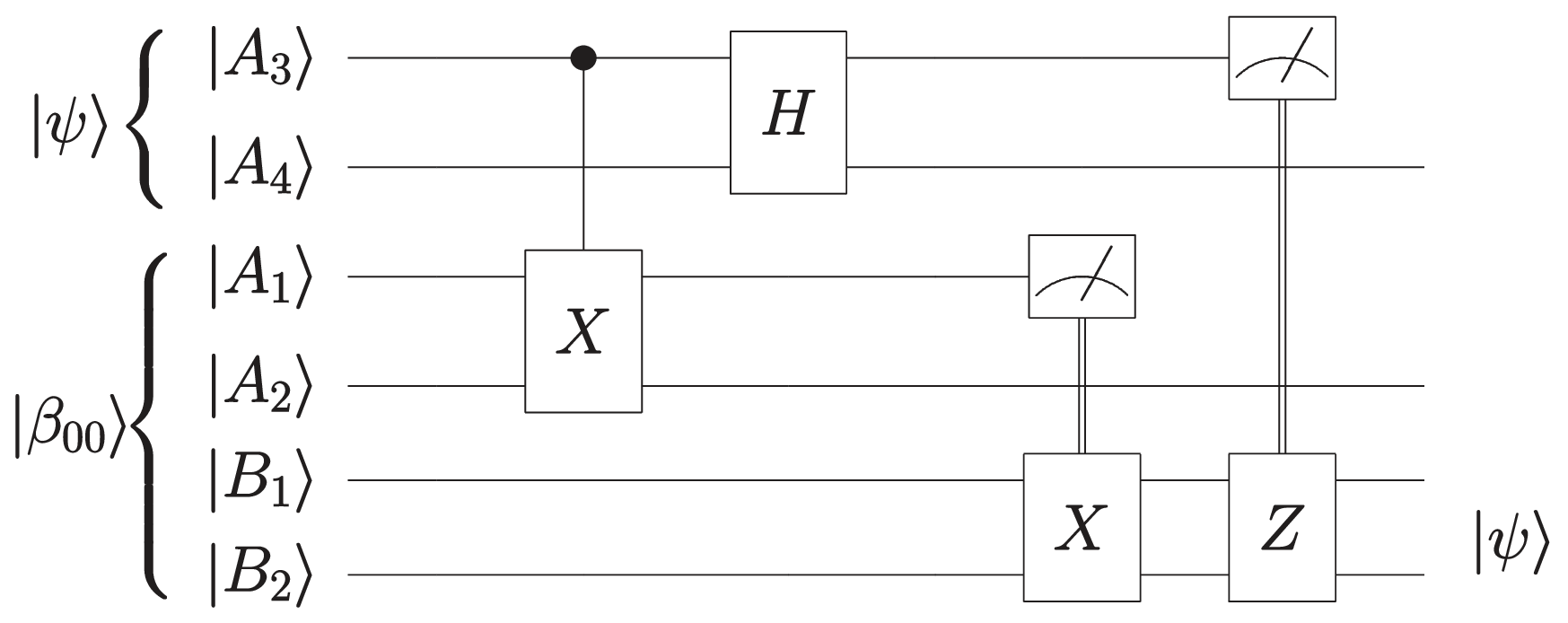}}}
\caption{Teleportation protocol with the present fermionic implementation. Each
qubit is represented by a pair of fermionic modes having a total occupation
number of 1. The control operation can be realized involving just one of the
modes of the pair representing the control qubit due to the occupation number
constraint, and similarly the usual measurement in the standard basis can be
implemented by measuring just one of these modes. If the pair occupation number
constraint is relaxed so that both local number parities coexist, then control
and measurement operations involve both modes.} \label{fig2}
\end{figure}

As an example, in Fig.\ \ref{fig2} we show the teleportation protocol adapted
to be implemented using an entangled fermion state as resource, and a two mode
sate to be teleported. Alice has the modes $\{|A_1\rangle, |A_2\rangle,
|A_3\rangle, |A_4\rangle\}$ while Bob is in possession of $\{|B_1\rangle,
|B_2\rangle \}$. The first two modes of Alice are entangled with those of Bob,
being in the joint state
$|\beta_{00}\rangle=\frac{1}{\sqrt{2}}(c^\dagger_{A_1}c^\dagger_{B_1}+c^\dagger_{A_2}
c^\dagger_{B_2})|0\rangle$,
and the remaining modes of Alice are in the state $|\psi\rangle=(\alpha\,
c^\dagger_{A_3}+ \beta\, c^\dagger_{A_4})|0\rangle$, $|\alpha|^2+|\beta^2|=1$.
The input state is therefore
\[
|\psi_i\rangle=\frac{1}{\sqrt{2}}(\alpha\, c^\dagger_{A_3}+ \beta\, c^\dagger_{A_4})
(c^\dagger_{A_1}c^\dagger_{B_1}+c^\dagger_{A_2}c^\dagger_{B_2})|0\rangle
\]
and it is straightforward to see that the output state is
\begin{eqnarray}
|\psi_o\rangle&=&\frac{1}{2}[
c^\dagger_{A_4}c^\dagger_{A_2}(\alpha\, c^\dagger_{B_1}+\beta\, c^\dagger_{B_2})
+c^\dagger_{A_4}c^\dagger_{A_1}(\alpha\, c^\dagger_{B_2}+\beta\, c^\dagger_{B_1})+
\notag\\&\phantom{[}&
c^\dagger_{A_3}c^\dagger_{A_2}(-\alpha\, c^\dagger_{B_1}+\beta\, c^\dagger_{B_2})
+c^\dagger_{A_3}c^\dagger_{A_1}(-\alpha\, c^\dagger_{B_2}+\beta\, c^\dagger_{B_1})]
|0\rangle\notag\nonumber
\end{eqnarray}
The controlled operations on Bob's modes depicted in Fig.\ \ref{fig2} then
ensure that his output will be the state $|\psi\rangle$.

Considering now a general circuit, if the input states are restricted to be
SD's in the previous basis, with one fermion for each pair, we recover a
classical circuit. The CNOT gate in (\ref{cnot}) reduces for these states to a
classical {\it controlled swap} or Fredkin gate, which implies that  reversible
classical computation can be done with SD's as input states.

On the other hand, if the occupation number restriction $N_{ij}=1$ (i.e., odd
number parity for each pair)  is relaxed, so that the building blocks of the
circuit are no longer single fermions  that can be found in two possible
states, but rather the fermionic modes themselves, other possibilities arise.
For instance, if now the input states contain either 0 or two fermions for each
pair (even  number parity qubits), such that modes $i,j$ are either both empty
or both occupied, then we should use the $\tilde{\sigma}_\mu^{ij}$ operators as
defined in (\ref{ssp2}), i.e.,
$\tilde{\sigma}^{ij}_x=c^\dagger_ic^\dagger_j+c_jc_i$,
$\tilde{\sigma}^{ij}_y=i(c_jc_i-c^\dagger_ic^\dagger_j)$,
$\tilde{\sigma}^{ij}_z=c^\dagger_ic_i+c^\dagger_jc_j-1$. In this  case the
operator $\tilde{U}^f_{_{\rm CNOT}}$ should be constructed as in Eq.\ (\ref{cnot})
with the $\tilde{\sigma}_\mu$ operators, while the operator (\ref{cnot}), and
in fact any unitary gate built with the $\sigma^{ij}_\mu$ operators,  will
become an identity for these states, as previously stated. Hence, by adding the
appropriate gates, the same modes can in principle be used for even and odd
number parity qubits independently.

For example, in the even local parity setting the input state for the teleportation
protocol would be 
\[
|\tilde{\psi}_i\rangle=\frac{1}{\sqrt{2}}(\beta+\alpha \, c^\dagger_{A_3}c^\dagger_{A_4})
(1+ c^\dagger_{A_1}c^\dagger_{A_2}c^\dagger_{B_1}c^\dagger_{B_2})|0\rangle\,.
\]
If $|0\rangle$ stands for a reference SD (Fermi sea), then this state involves
$0$, one, two and three particle hole excitations, with  $A_4,A_2,B_2$,
standing for holes. The output state becomes
\begin{eqnarray}
|\tilde{\psi}_o\rangle&=&{\textstyle\frac{1}{2}}[
(\beta+\alpha\, c^\dagger_{B_1}c^\dagger_{B_2})
+c^\dagger_{A_1}c^\dagger_{A_2}(\alpha+\beta\,c^\dagger_{B_1}c^\dagger_{B_2})
+c^\dagger_{A_3}c^\dagger_{A_4}\times\notag\\&\phantom{[}&(\beta-\alpha\, 
c^\dagger_{B_1}c^\dagger_{B_2})
+c^\dagger_{A_1}c^\dagger_{A_2}c^\dagger_{A_3}c^\dagger_{A_4}(-\alpha+\beta\, 
c^\dagger_{B_1}c^\dagger_{B_2})]
|0\rangle\notag\nonumber\,,
\end{eqnarray}
so that if Alice measures which of her modes are occupied and sends  the result to Bob, 
he can reconstruct the original state by applying the pertinent 
$\tilde{X}\equiv i e^{-i\frac{\pi}{2}\tilde{\sigma}^{12}_x}$ and
$\tilde{Z}\equiv i e^{-i\frac{\pi}{2}\tilde{\sigma}^{12}_z}$ operators.

Finally, let us consider the case of  {\it superdense coding} \cite{SD.96,NC.00}. 
It is clear from the previous discussion that it can be implemented with the fermionic
$|\beta_{00}\rangle$ state of the teleportation example and performing exactly
the  same local operations of the usual case, but viewed now as two-mode
operations. Now a general state with even global parity of the four modes
$\{|A_1\rangle, |A_2\rangle, |B_1\rangle, |B_2\rangle\}$ is a combination  of
eight states as in Eq.\ (\ref{eq1}): six two-particle states, the vacuum
$|0\rangle$ and the completely occupied state $|\bar 0\rangle$, as shown in 
Fig.\ \ref{fig1}. Four of the six two-particle states (top of Fig.\ \ref{fig1}), 
have  $N_{A}=N_{B}=1$ and can be used to  reproduce the known results of the standard 
protocol. But the four remaining states, which have even local parity, may be used as 
well for superdense coding if the proper local operations expressed in terms of the 
${\tilde\sigma^{AB}_\mu}$ are performed. 

A general even parity state (\ref{eq1}) may then be thought of as a
superposition of states of two different two-qubit systems, like in Eqs.\ 
(\ref{Psi}) and (\ref{st6}). Defining the maximally entangled orthogonal definite 
local parity states 
\begin{eqnarray}
\!\!\!|\beta_{00\atop 10}\rangle&=&\gamma(c^\dagger_{A_1}c^\dagger_{B_1}\pm
c^\dagger_{A_2}c^\dagger_{B_2})|0\rangle,
|\tilde\beta_{00\atop 10}\rangle=\gamma(c^\dagger_{A_1}c^\dagger_{A_2}
c^\dagger_{B_1}c^\dagger_{B_2}\pm 1)|0\rangle\nonumber\\
\!\!\!|\beta_{01\atop 11}\rangle&=&\gamma(c^\dagger_{A_1}c^\dagger_{B_2}
\pm c^\dagger_{A_2}c^\dagger_{B_1})|0\rangle,
|\tilde\beta_{01\atop 11}\rangle=\gamma(c^\dagger_{A_1}c^\dagger_{A_2}
\pm c^\dagger_{B_1}c^\dagger_{B_2})|0\rangle\nonumber
\end{eqnarray}
with $\gamma=\frac{1}{\sqrt{2}}$, we may consider for instance the state
\begin{equation}
|\Psi_{00}\rangle={\textstyle\frac{1}{\sqrt{2}}}(|\beta_{00}\rangle
+|\tilde\beta_{00}\rangle)\,.\label{st8}
\end{equation}
By implementing on (\ref{st8}) the identity and the local operations 
$ie^{-i\frac{\pi}{2}(\sigma^A_\mu+\tilde\sigma^A_\mu)}=
 \sigma_\mu+\tilde\sigma_\mu$, $\mu=x,y,z$, and taking into account Eq.\ (\ref{kern}), 
 Alice can generate four orthogonal states: $|\Psi_{00}\rangle$ and 
\begin{subequations}
\begin{eqnarray}
\!\!\!\!\!\!|\Psi_{01}\rangle&=&ie^{-i\frac{\pi}{2}(\sigma^A_x+\tilde\sigma^A_x)}
|\Psi_{00}\rangle={\textstyle\frac{1}{\sqrt{2}}}(|\beta_{01}\rangle+|\tilde\beta_{01}\rangle)\,,\;\\
\!\!\!\!\!|\Psi_{10}\rangle&=&ie^{-i\frac{\pi}{2}(\sigma^A_z+\tilde\sigma^A_z})|\Psi_{00}\rangle
={\textstyle\frac{1}{\sqrt{2}}}(|\beta_{10}\rangle+|\tilde\beta_{10}\rangle)\,,\;\\
\!\!\!\!\!\!|\Psi_{11}\rangle&=&-e^{-i\frac{\pi}{2}(\sigma^A_y+\tilde\sigma^A_y)}
|\Psi_{00}\rangle
={\textstyle\frac{1}{\sqrt{2}}}(|\beta_{11}\rangle+|\tilde\beta_{11}\rangle)\,.\;
\end{eqnarray}
\label{dc1}
\end{subequations}
But she  can also perform these operations with a local parity gate
$P^A=-\exp[i\pi N_A]$ that changes the sign of local even parity states. This
allows her to locally generate another set of four orthogonal states,
\begin{eqnarray}
\!\!\!\!\!\!|\tilde\Psi_{ij}\rangle&=&P^A|\Psi_{ij}\rangle={\textstyle\frac{1}{\sqrt{2}}}
(|\beta_{ij}\rangle-|\tilde\beta_{ij}\rangle)\,,\;\;i,j=0,1\,,
\label{dc2}
\end{eqnarray}
which are orthogonal to each other and to the states (\ref{st8})--(\ref{dc1}). Hence, 
by relaxing the occupation number constraint on the partitions  it is possible for Alice 
to send  8 orthogonal states to Bob, i.e. three bits of information, using only two modes 
and local unitary operations that preserve the local parity, while with one type of qubits and 
the same operations she can send only two bits. Of course, if  parity restrictions were absent 
and she could change the local (and hence the global) parity she  could send four bits (in agreement 
with the maximum capacity for two  $d=4$ qudits, which is  $\log_2 d^2$ \cite{SD.96}). A fixed global 
parity constraint reduces the total number of orthogonal states she can send to Bob by half.   

On the other hand, since the state (\ref{st8}) does not have a definite local number parity, 
the ensuing bipartite entanglement is not restricted by the fermionic entanglement as shown 
in sec.\ \ref{F}. In fact all previous 8 states (\ref{st8}), (\ref{dc1}) and (\ref{dc2}) have 
maximum bipartite entanglememt, leading to maximally mixed reduced states $\rho_{A(B)}$: 
$S(\rho_A)=S(\rho_B)=2$, 
while by applying Eq.\ (\ref{Cex1}) it is seen that the fermionic concurrence of
the previous states is $C(|\Psi_{ij}\rangle)=C(|\tilde{\Psi}_{ij}\rangle)=1$.
The unitary operations applied by Alice are local and hence cannot change the bipartite
entanglement, while they are also one-body unitaries (i.e., exponents of quadratic fermion operators) 
so that they cannot change the fermionic concurrence and entanglement (i.e., the eigenvalues of 
$\rho^{\rm qsp}$) either. In fact, the fermionic entanglement is here not required. By changing the 
seed state (i.e., $|\Psi'_{00}\rangle=\frac{1}{\sqrt{2}}(|\beta_{00}\rangle+|\tilde{\beta}_{10}\rangle)$),
it is possible for Alice to generate locally 8 orthogonal states with the same
bipartite entanglement yet no fermion entanglement ($C(|\Psi'_{00}\rangle)=0$).

Therefore the entanglement built with local states with different number parity
plays the role of a resource for superdense coding. In fact even the state 
(\ref{psis}) with $\alpha=\beta=\frac{1}{\sqrt{2}}$, which has obviously null concurrence,
 can in principle be used for sending two bits if Alice can perform the parity preserving operations 
$P^A=-\exp[i\pi N_A]$, $\sigma_x+\tilde\sigma_x$ and
$P^A(\sigma_x+\tilde\sigma_x)$.  
It is worth noting, however, that the same state cannot be directly used as a resource
for teleportation with the standard protocol without violating the parity superselection rule, 
since Bob's two local states have opposite parity and cannot be
superposed. After a measurement of Alice's modes Bob's reduced
state will collapse to a state of definite parity in a realizable protocol, so that it will be
impossible for him to recover a general state $|\psi\rangle$.  

We have so far considered just the number parity restriction. If other superselection rules 
(like charge or fermion number) also apply for a particular realization they will imply stronger 
limitations on the capacity of states like (\ref{st8}). Nonetheless, even local parity qubits 
with no fixed fermion number remain realizable through particle-hole realizations, i.e., 
excitations over a reference Fermi sea in a many-fermion system. 
 
We also mention that a basic realization of four dimensional sp space-based fermionic qubits 
is that of a pair of spin $1/2$ fermions in the two lowest states of a double well scenario 
in a magnetic field, which would control the energy gap between both spin directions and the 
transitions between them. For single occupation of each well we would have odd local parity 
qubits, while allowing double or zero occupancy through hopping between wells we could also 
have even local parity qubits. 

\section{Conclusions\label{IV}}

We have first shown that there is an exact correspondence between bipartite
states and two-fermion states of the form (\ref{st4}) having a  fixed local
number parity. Entangled states are represented by fermionic states which are
not Slater Determinants, and reduced local states correspond to blocks of the 
sp density matrix. In particular, qubits can be represented by pairs of
fermionic modes with occupation number restricted to $1$ (odd number parity qubits). 
This result allows to rewrite qubit-based quantum circuits as fermionic circuits. 
But in addition, a fermionic system also enables zero or double occupancy of 
these pairs, which gives rise to a second type of qubit (even number parity qubits).  
Dual type circuits can then be devised, as the gates for one parity become 
identities for the other parity. And even though both types of qubits cannot 
be locally superposed due to the parity superselection rule, they can contribute 
to the entanglement in a global fixed parity state. 

We have then demonstrated rigorous properties of the basic but fundamental case 
of a four-dmensional sp space. First, there is always a single-particle (or quasiparticle) 
basis in which any pure	state can be seen as a state of two distinguishable qubits, 
with  the fermionic concurrence determining the entanglement between these two qubits 
(Eq.\ (\ref{Sfq})). Such entanglement is ``genuine'', in the sense that the local states 
involved have a definite parity and can therefore be combined. Secondly, such fermionic 
entanglement was shown to provide  always a lower bound to the entanglement obtained 
with  any other bipartition of this sp space, although the extra entanglement arises 
from the superposition of states with different local parity. While its capacity for 
protocols involving superpositions of local states is limited, such entanglement 
can nevertheless still be useful for other tasks such as superdense coding. 

\acknowledgments
The authors acknowledge support from CONICET (N.G.)
and CIC (R.R.) of Argentina. Work supported by CIC and
CONICET PIP 112201501-00732.

\end{document}